\documentstyle[epsf,preprint,aps]{revtex}
\draft
\preprint{SNUTP 99/051}
\begin{document}
\title{\Large\bf Effective Gauss-Bonnet Interaction 
in Randall-Sundrum Compactification 
}
\author{Jihn E. Kim, Bumseok Kyae and Hyun Min Lee} 
\address{ Department of Physics and Center for Theoretical
Physics, Seoul National University,
Seoul 151-742, Korea}
\maketitle

\begin{abstract} 
The effective gravitational interaction below the Planck scale in the
Randall-Sundrum world is shown to be the Gauss-Bonnet term. In this
theory we find that there exists another static solution with a 
positive bulk cosmological constant. Also, there
exist solutions for positive visible sector cosmological constant,
which is needed for a later Friedman-Robertson-Walker universe.
\end{abstract}
\pacs{11.25.Mj, 12.10.Dm, 98.80.Cq}

\newpage

In the last several years, the brane world has been one
of the major research fronts toward probing the fundamental laws
of nature \cite{hw,dd,rs}. In many of these theories, the observable
matter fields live in the brane(s) and are forbidden to propagate in the
bulk. The bulk is populated mainly by the gravity related fields
\cite{dd,rs}. Among these, the Randall-Sundrum(RS) compactification
touches upon the gauge hierarchy problem($\lq\lq$Why is $M_W/M_P$ 
so small?), and obtains an exponentially small Higgs boson mass parameter
compared to the input mass parameter(assumed to be at the Planck 
scale) of the five dimensional theory. 
An exponential warp factor
suppresses the soft mass in the visible brane, and it is possible to
obtain this small ratio because the Higgs mass term is a
dimension two operator. Thus, 
in the Randall-Sundrum(RS) world, one
changes the traditional gauge hierarchy problem to a problem
in geometry.

However, the RS model is not complete toward a solution of the gauge
hierarchy problem. The most obvious problem is
the appearance of unwanted high dimensional 
operators such as the one triggering proton decay. Generally, the
TeV scale is the mass parameter scale suppressing the high dimensional 
($d>4$) operators in the RS compactification. Thus one has to 
make sure that the theory has a high degree of symmetry to suppress
the unwanted operators up to the experimental limit. Another problem is
the problem of inflation. Generally, the unwanted inflation
is unavoidable unless one fine-tune the bulk cosmological constant 
and the brane tensions \cite{kal,nihei,kimkim}. 
Nevertheless, this scenario is
so interesting that it is worthwhile to pursue the implications of
the RS compactification further.
The most interesting point of the RS world is the interplay of the
bulk and the brane world. In particular, the bulk cosmological
constant ($k$) and the brane tensions ($k_i\ (i=1,2)$) must be 
related, $k_1=-k=-k_2$ and $k_1>0$. Note, however, that the observed
expansion rate of the observable universe is measured by the Hubble parameter
which is vanishing, given the above condition, neglecting
the matter field Lagrangian.  
These $k$'s are the appropriately
defined from the original bulk cosmological constant $\Lambda_b$ and the
brane tensions $\Lambda_1, \Lambda_2$ of the two branes 1 and 2 where
the hidden fields live in Brane 1 (B1) and the observable fields live in 
Brane 2 (B2) \cite{inflation}. 
For simplicity, we will call these $\Lambda$'s the $\lq$cosmological 
constants', below. Therefore, we can hope for a possibility of understanding 
the old cosmological costant problem in this new setting. 

In this paper, we consider the leading effective interactions
below the Planck scale with the setting of
two branes located at $y=0$ and $y=1/2$ and with an
orbifold symmetry $S_1/Z_2$. We find that two static 
solutions exist only if the effective interaction is the
Gauss-Bonnet term. One of these solutions reduces to
the RS one in the limit of the vanishing effective interaction
with a negative cosmological constant in the observable brane.
Another solution is a new one with the possibility of 
positive cosmological constants in the observable brane.

We proceed to discuss the effective interaction
and its solution with an $S_1/Z_2$ orbifold compactification.
The space-time is assumed to be five dimensional ($M, N = 0, 1, \cdots,
4$) with the four dimensional brane worlds ($\mu, \nu=0, 1, \cdots, 3; 
y=0,1/2$) and the bulk ($M,N=0,\cdots,4;0< y<1/2$) \cite{rs},
where $y=x^4$. The $S_1/Z_2$ orbifold is used to
locate the two branes at $y=0$ and $y=1/2$. Thus, the periodicity is
$y\rightarrow y+1$.

Below the Planck scale, the gravity effects can be added as 
additional effective interaction terms. Since we are neglecting the
matter interactions, the possible terms in the Lagrangian is, up to 
O($g_{MN}^2$)
$$
S=\int d^5x\sqrt{-g}\left( {M^3\over 2}R-\Lambda_b +\frac{1}{2}\alpha M R^2
+\frac{1}{2}\beta M R_{MN}R^{MN}+\frac{1}{2}\gamma MR_{MNPQ}R^{MNPQ}\right)
$$
\begin{equation} 
+\sum_{i=1,2\ {\rm branes}}\int d^4x\sqrt{-g^{(i)}}\left({\cal L}_i
- \Lambda_i\right)
\end{equation}  
where $g, g^{(i)}$ are the determinants of the metrics in the
bulk and the branes, $M$ is the five dimensional gravitational 
constant, $\Lambda_b$
and $\Lambda_i$ are the bulk and brane cosmological constants,
and $\alpha,\beta,\gamma$ are the effective couplings.
We assume that the three dimensional space is homogeneous and 
isotropic, and hence the metric is parametrized by $n, a$, and $b$
\begin{equation}
ds^2=-n^2(\tau,y)d\tau^2+a^2(\tau,y)\delta_{ij}dx^idx^j+b^2(\tau,y)dy^2.
\end{equation}
where the lower case Roman characters $i,j$ denotes the space 
indices 1, 2, and 3.

Then static solutions exist only when the Gauss-Bonnet relations
$\beta=-4\alpha$ and $\gamma=\alpha$ are satisfied.
Namely, the higher $y$ derivative terms, 
$a^{\prime\prime\prime\prime}$,
$a^{\prime\prime\prime}a^\prime$, and $a^{\prime\prime}a^{\prime\prime}$
(where prime denotes the derivative with respect to
$y$) are absent in the l.h.s. of the Einstein equation.
This condition is necessary since the r.h.s. of the
Einstein equation contains only
one power of the Dirac delta function, and the higher derivative
terms diverge more rapidly than the delta function at the
branes.  {\it It is miraculous that there
exists such a solution}, since the number(=3) of conditions we must satisfy 
are more than the number(=2) of independent ratios of the couplings
$\alpha,\beta$ and $\gamma$. 
Thus, we use the Gauss-Bonnet effective interaction below.
This condition leads to ghost-free nontrivial gravitational
self interactions for dimensions higher than four \cite{ghost}. 

Variations of the above action with the Gauss-Bonnet 
term gives, apart from those for the brane Lagrangian,
$$
\sqrt{-g}\bigg[R_{MN}-{1\over 2}g_{MN}R-{1\over 2}\alpha M^{-2}g_{MN}\left(
R^2-4R_{PQ}R^{PQ}+R_{STPQ}R^{STPQ}\right)
$$
$$
+2\alpha M^{-2}\left(RR_{MN}-4R_{MP}R_N\,^P+R_{MQSP}R_N\,^{QSP}\right)
+2\alpha M^{-2}\left(g_{MN}R_{;P}\,^{;P}-R_{;M;N}\right)
$$
\begin{equation}
-4\alpha M^{-2}\left(g_{MN}R^{PQ}\,_{;P;Q}+R_{MN;P}\,^{;P}
-R_M\,^P\,_{;N;P}-R_N\,^P\,_{;M;P}\right)
\end{equation}
$$
+2\alpha M^{-2}\left(R_M\,^P\,_N\, ^Q\,_{;P;Q}-R_M\,^P\,_N\,^Q 
\,_{;Q;P}\right)\bigg]
$$
$$
=-M^{-3}\left[\Lambda_b\sqrt{-g}g_{MN}+\Lambda_1\sqrt{-g^{(1)}}
g^{(1)}_{\mu\nu}\delta^\mu_M\delta^\nu_N\delta(y)+\Lambda_2\sqrt{-g^{(2)}}
g^{(2)}_{\mu\nu}\delta^\mu_M\delta^\nu_N\delta(y-\frac{1}{2})\right]
$$
where 1 refers to the brane of the hidden world B1 and 2 refers to
the observable brane B2. The l.h.s. of the above equation 
contains the extra term due to the Gauss-Bonnet term,
$\sqrt{-g}X_{MN}$, in addition to the familiar Einstein 
tensor $\sqrt{-g}G_{MN}$. The $G_{MN}$ and $X_{MN}$ are
\begin{equation}
G_{00}= 
-3\left({n^2{a^{\prime\prime}}\over {ab^2}}
+{n^2{a^{\prime 2}}\over {a^2b^2}}
-{n^2{a^\prime b^\prime}\over {ab^3}}\right)
+\cdots
\end{equation}
\begin{equation}
G_{ij}= 
\left({{n^{\prime\prime}}a^2\over {nb^2}}+2{a{a^{\prime\prime}}\over {b^2}}
+2{{n^\prime aa^\prime}\over {nb^2}}-{{n^\prime a^2  b^\prime}\over {nb^3}}
+{{a^{\prime 2}}\over {b^2}}-2{a{a^\prime b^\prime}\over {b^3}}\right)
\delta_{ij}+\cdots
\end{equation}
\begin{equation}
G_{55}= 
3\left({{n^\prime a^\prime}\over {na}}
+{{a^{\prime 2}}\over {a^2}}\right) 
+\cdots
\end{equation}
\begin{equation}
G_{05}= 
\cdots
\end{equation}
\begin{equation}
X_{00}=12\alpha \left({a^{\prime 2}a^{\prime\prime}n^2\over a^3 b^4}
-{a^{\prime 3}b^\prime n^2\over a^3b^5}\right)+\cdots
\end{equation}
\begin{equation}
X_{ii}=-4\alpha\left(2{a^\prime a^{\prime\prime}n^\prime\over nb^4}
+{n^{\prime\prime}a^{\prime 2}\over nb^4}
-3{a^{\prime 2}n^\prime b^\prime\over nb^5}\right)+\cdots
\end{equation}
\begin{equation}
X_{55}=-12\alpha{a^{\prime 3}n^\prime\over a^3nb^2}+\cdots
\end{equation}
\begin{equation}
X_{05}=\cdots
\end{equation}
where $\prime$ denotes the derivative with repect to $y$ and 
$\cdots$ denote terms containing derivatives with respect to $\tau$
\cite{kkl}.

To find static solutions, let us redefine $y$ such that the metric
takes the following form,
\begin{equation}
ds^2=e^{-2\sigma(\phi)}\eta_{\mu\nu}dx^\mu dx^\nu+r_c^2 d\phi^2
\end{equation}
where the length parameter $r_c$ is a constant, and the angle variable
is defined as $\phi=2\pi y$, with the
periodicity $\phi\rightarrow\phi+2\pi$. 
Note that the Einstein equation for the (00) 
component is identical to the $(ii)$ component. Thus the (00), $(ii)$,
and (55) components of Eq.~(3) lead to
\begin{equation}
{3\sigma^{\prime\prime}\over r_c^2}\left(1-{4\alpha\over M^2r_c^2}
(\sigma^\prime)^2\right)=
{\Lambda_1\over M^3 r_c}\delta(\phi)+{\Lambda_2\over 
M^3 r_c}\delta(\phi-\pi)
\end{equation}
\begin{equation}
{6(\sigma^\prime)^2\over r_c^2}\left(
1-{2\alpha\over M^2 r_c^2}(\sigma^\prime)^2\right)={-\Lambda_b\over M^3}
\end{equation} 

There exist two solutions of Eq.~(14), consistent with the
orbifold symmetry $\phi\rightarrow -\phi$,
\begin{equation}
\sigma^{\pm}=r_c|\phi|\left[{M^2\over 4\alpha}\left(
1\pm \left(1+{4\alpha\Lambda_b\over 3M^5}\right)^{1\over 2}\right)\right]
^{1\over 2}\equiv k_{\pm}r_c|\phi|.
\end{equation}
These solutions exist for:\\  
\indent (i) $\alpha<0$ and $\Lambda_b<0$ allows only $\sigma^-$, and\\
\indent (ii) $\alpha>0$ allows both $\sigma^\pm$ solutions. The $\sigma^+$ 
solution is possible for both $\Lambda_b>0$ and\\
\indent $\Lambda_b<0$.  The
$\sigma^-$ solution is possible only for $\Lambda_b<0$.
In any case, there exists the lower 
\indent limit of $\alpha\Lambda_b$,
$\alpha\Lambda_b\ge -3M^5/4$. 

Considering the discontinuities at the branes given in Eq.~(13),
we obtain two solutions if the following relations among the brane
cosmological constants are satisfied
$$
\Lambda_1^{\mp}=-\Lambda_2^{\pm}=\mp6k_{\pm}M^3
\sqrt{1+\frac{4\alpha\Lambda_b}{3M^5}}
$$
\begin{equation}
=\mp 6M^3\left[{M^2\over 4\alpha}\left(1\pm \left(
1+{4\alpha\Lambda_b\over 3M^5}\right)^{1\over 2}\right)
\left(1+{4\alpha\Lambda_b\over 3M^5}\right)\right]^{1\over 2}
\end{equation}
where $k_\pm>0$. The RS solution is obtained by taking
$\alpha\rightarrow 0$ in the -- solution. 

Possible solutions are depicted in Fig.~1 as a function of the
Gauss-Bonnet coupling $\alpha$. The vertical axis$(\equiv \lambda_2)$ 
is the solution for $\Lambda_2$ in the visible brane in units of
$\sqrt{6M^3|\Lambda_b|}$, and the horizontal axis $(\equiv \alpha_\Lambda)
$ is defined as $4\alpha\Lambda_b/(3M^5)$.

In the literature, it has been argued that $\Lambda_{\rm vis}
=\Lambda_2$ is better to be positive \cite{hubble}. 
The argument is the following.  The hubble parameter is expressed in
terms of the cosmological constant and the energy density,
$H_{\pm,\rm vis}=\sqrt{ (k_{\pm,\rm vis})^2
-k^2}$ \cite{kkl}, where $k^2=k^2_\pm$ for the static
solutions and the two parameters corresponding to + and -- solutions 
at the visible brane, $k_{\pm,\rm vis}$, are given by
\begin{equation} 
k_{\pm,\rm vis}={\pm \left(\Lambda_2^\pm+\rho_{\rm vis}\right)\over
6M^3\sqrt{1+({4\alpha\Lambda_b}/{3M^5})}}
\end{equation}
Thus the Hubble parameter at B2 is given by
\begin{equation}
H_{\pm,\rm vis}=\frac{\sqrt{\rho_{\rm vis}(\rho_{\rm vis}+2\Lambda_2^\pm)
}}{6M^3\sqrt{1+(4\alpha\Lambda_b/3M^5)}}
\end{equation}
With $\rho_{\rm vis}=0$, we obtain the previous 
static solution. But with $\Lambda^\pm_2=\Lambda_b<0$, there exists a
possibility that $\rho_{\rm vis}(2\Lambda_2^\pm+\rho_{\rm vis})<0$ 
at a sufficiently low temperature, and hence it is
difficult to obtain a real Hubble parameter \cite{hubble}. 
But with a positive
$\Lambda_2^+$, there does not exist such a problem.
This is possible for our + solution for $\alpha>0$. 

Inflationary solutions were obtained for a flat bulk geometry
\cite{lukas} and for an AdS bulk geometry \cite{kal,nihei}.
In our case, inflationary solutions exist also for a positive
bulk cosmological constant \cite{kkl}.
In general, inflation occurs if parameters $\alpha,\Lambda_b,\Lambda_1$,
and $\Lambda_2$ do not satisfy the two relations implied by
Eq.~(16) \cite{nihei,kimkim,kkl}.  We find that
in general there exist terms with higher (more than two) time derivatives
if the parameters of Eq.~(1) satisfy $16\alpha+5\beta+4\gamma=0$. 
But the inflationary solutions in Ref.~\cite{nihei} (with a separable
metric) and Ref.~\cite{kimkim} (with a nonseparable metric) 
are still valid with these higher time derivatives if $16\alpha+5\beta
+4\gamma=0$ is satisfied \cite{kkl}.

The Planck constant at B2 (observable sector) is given by \cite{rs}
$$
{M_{Pl}}^2=M^3r_c\int_{-\pi}^{\pi}d\phi 
e^{-2k_{\pm}r_c|\phi|}=\frac{M^3}{k_{\pm}}[1-e^{-2k_{\pm}r_c\pi}]
$$
\vskip -0.5cm
\begin{equation}
=M^2\bigg[\frac{1}{4\alpha}\bigg(1
\pm\Big(1+\frac{4\alpha \Lambda_{b}}{3M^5}\Big)^{\frac{1}{2}}
\bigg)\bigg]^{-\frac{1}{2}}[1-e^{-2k_{\pm}r_c\pi}]
\end{equation}
The Higgs boson mass parameter at the observable sector is obtained
by redefining the Higgs field such that the kinetic energy
term of the Higgs boson takes a standard form \cite{rs}. Thus the Higgs
mass parameter is given by
\begin{eqnarray}
m&\equiv& e^{-k_{\pm}r_c\pi}m_0\nonumber\\
&=&m_0\, \exp\bigg(-r_c\pi\bigg[\frac{M^2}
{4\alpha}\bigg(1\pm\Big(1+\frac{4\alpha 
\Lambda_{b}}{3M^5}\Big)^{\frac{1}{2}}\bigg)\bigg]^{\frac{1}{2}}\bigg)
\end{eqnarray}
where $m_0$ is the mass given in the fundamental Lagrangian, before
redefining the Higgs field.
For $k_+r_c\simeq 12$, the + solution gives a needed large
mass hierarchy through the warp factor $e^{-k_+r_c\pi}$
from the input mass parameter($M$)  of order
$10^{19}$~GeV, leading to a TeV scale observable mass. For
$k_-r_c\simeq 12$, the -- solution has the same behavior.
This small warp factor \cite{rs} makes it possible to generate
a TeV scale mass from the fundamental parameter of
O($M_{Pl}$). But these TeV scale masses also appear in the other mass
parameters of the effective operators, in particular those
leading to proton decay are also parametrized by a TeV scale
mass. Therefore, one has to suppress sufficiently the low
dimensional proton decay operators such that it is sufficiently 
long-lived ($\tau_p>10^{32}$~years), implying $D>14$.

In this paper, we studied the Randall-Sundrum compactification with
an additional effective Gauss-Bonnet term with the coupling
$\alpha$. It is miraculous that
the Gauss-Bonnet term is allowable for a static solution, since
it is not at all obvious to expect such an effective Lagrangian.
This is because the conditions we must satisfy are more than
the independent ratios of the effective couplings. With this 
effective Gauss-Bonnet term, we find two solutions. There exist
solutions for negative and positive cosmological constants in
the visible sector. In particular, we obtained a solution for 
a positive cosmological constant in the visible sector in case
$\alpha>0$, which can introduce a positive matter energy density
in the visible Friedman-Robertson-Walker universe.

\acknowledgments
This work is supported in part by the BK21 program of Ministry 
of Education, and by the Korea Science and Engineering Foundation
(KOSEF 1999 G 0100).

\newpage 
\begin{figure}
\epsfxsize=160mm
\centerline{\epsfbox{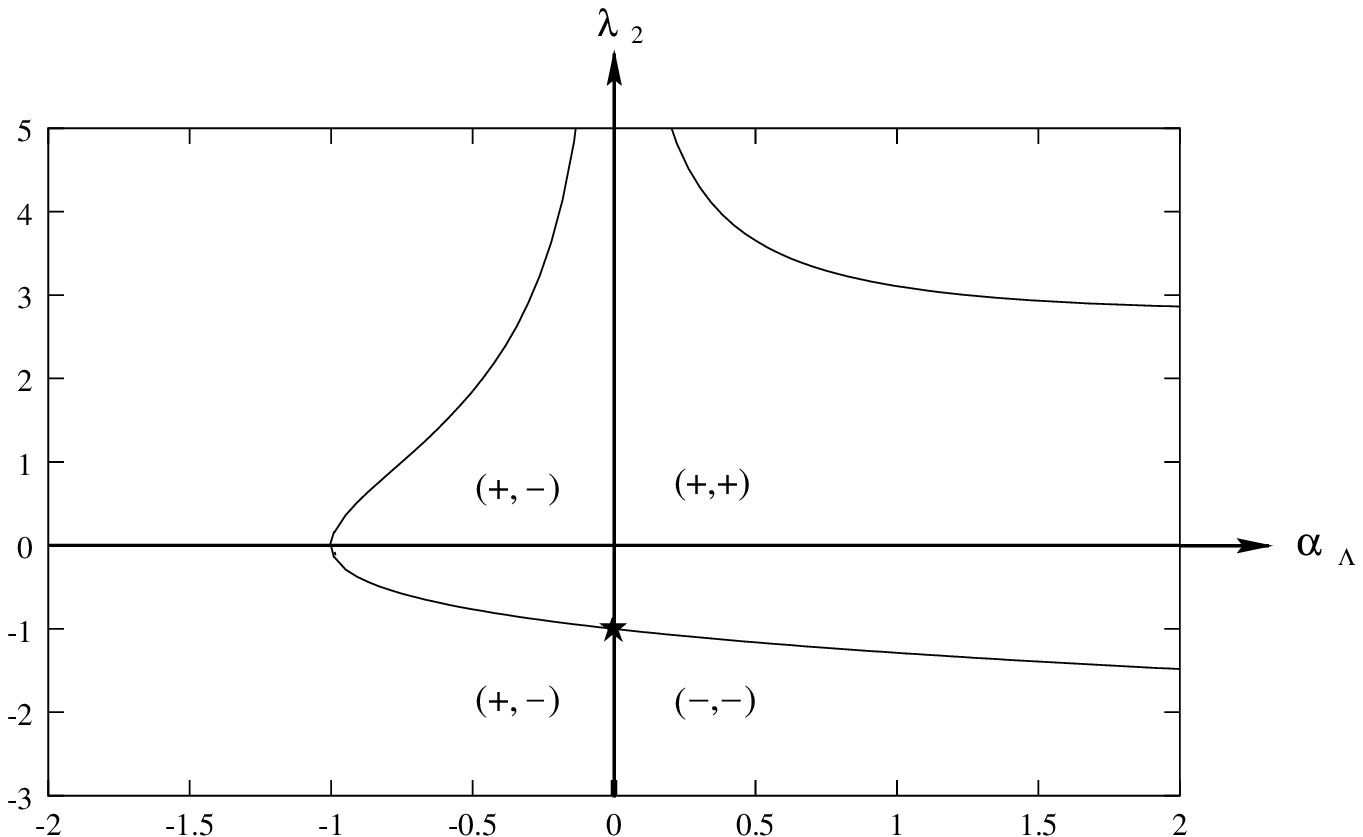}}
\end{figure}
\noindent Fig.~1. Possible solutions for $\lambda_2\equiv
\Lambda_2/\sqrt{6M^3|\Lambda_b|}$  as a function of 
$\alpha_{\Lambda}\equiv 4\alpha\Lambda_b/(3M^5)$. The star point
is the RS solution. The four quadrants have different sets of
signs of $\alpha$ and $\Lambda_b$, denoted as $({\rm sign\ of\ }
\alpha,\ {\rm sign\ of\ } \Lambda_b)$. 


\begin{references}
\bibitem{inflation} Notations for $k$'s are borrowed from
Refs.~\cite{nihei,kimkim}.
\end{references}
\end{document}